\documentclass[conference]{IEEEtran}
\IEEEoverridecommandlockouts

\usepackage{etex}
\usepackage{amsmath,amssymb}
\usepackage{float}
\usepackage{lipsum}
\usepackage{array}
\usepackage{cite}
\usepackage{textcomp}
\usepackage{hhline}
\usepackage{siunitx}
\usepackage{balance}
\usepackage{makecell}
\usepackage{float}
\usepackage[pdftex]{graphicx}
\usepackage{siunitx}
\usepackage[dvipsnames]{xcolor}
\usepackage{xcolor}
\usepackage{colortbl}
\usepackage{tikz}
\usetikzlibrary{decorations}
\usetikzlibrary{calc} 
\tikzset{>=latex}
\usetikzlibrary{arrows,shapes}
\usetikzlibrary{arrows}
\usetikzlibrary{plotmarks}
\usepackage{ctable}
\usepackage{pgfplots}
\usepackage[english]{babel}
\usepackage{color}
\usepackage[utf8]{inputenc}
\usepackage[T1]{fontenc}
\usetikzlibrary{decorations.pathreplacing,shapes.misc}
\usetikzlibrary{patterns}
\usepackage{tcolorbox}
\usepackage{balance}
\usepackage{subfigure}
\usepackage{hyphenat}

\addto\captionsenglish{\renewcommand{\figurename}{Fig.}}

\newcommand{\RF}{}

\renewcommand{\a}{\boldsymbol{a}}
\renewcommand{\b}{\boldsymbol{b}}

\newcommand{\nc}{n}

\newcommand{\kc}{k}

\newcommand{\lr}{\boldsymbol{r}}

\newcommand{\lrt}{\tilde{\boldsymbol{r}}}
\newcommand{\lc}{\boldsymbol{c}}
\newcommand{\llr}{\boldsymbol{l}}
\newcommand{\rr}{\boldsymbol{R}}
\newcommand{\rbol}{\boldsymbol{r}}

\newcommand{\cc}{\boldsymbol{C}}

\newcommand{\dgmd}{\mathsf{d}_\mathsf{GD}}
\newcommand{\dmin}{d_\mathsf{min}}
\newcommand{\dGMD}{\dgmd(\lr,\hat{\lr})}

\newcommand{\lalone}{\boldsymbol{L}}

\newcommand{\w}{w}
\newcommand{\BB}{\mathsf{B}}

\def\forcemath#1{\ifmmode #1 \else $#1$\fi}

\newcommand{\ham}{\mathsf{d}_\mathsf{H}}

\ifCLASSINFOpdf
 \else
  \fi

\begin{document}


\title{On Low-Complexity Decoding of Product Codes for High-Throughput Fiber-Optic Systems}

\author{Alireza Sheikh$^\mathsection$, Alexandre Graell i
Amat$^\mathsection$, Gianluigi Liva$^\dagger$, Christian
H{\"a}ger$^{\mathsection,\ddagger}$, and Henry D.~Pfister$^\ddagger$ \\ \IEEEauthorblockA{$^\mathsection$ Department of Electrical Engineering, Chalmers University of Technology, Sweden \\ $^\dagger$Institute of Communications and Navigation of the German Aerospace Center (DLR), Germany \\ $^\ddagger$ Department of Electrical and Computer Engineering, Duke University, Durham, NC, USA\\~\\ (Invited Paper)
\thanks{This work was financially supported by the Knut and Alice
Wallenberg and the Ericsson Research Foundations. This work is also part of a project that has received funding
	from the European Union's Horizon 2020 research and innovation
	programme under the Marie Sk\l{}odowska-Curie grant 
	No.~749798. The work was also supported by the National
	Science Foundation (NSF) under grant No.~1609327. Any opinions,
	findings, recommendations, and conclusions expressed in this
	material are those of the authors and do not necessarily reflect the
	views of these sponsors.}
}}


\maketitle


\begin{abstract}
We study low-complexity iterative decoding algorithms for product
codes. We revisit two algorithms recently proposed by the
authors based on bounded distance decoding (BDD) of the component
codes that improve the performance of conventional iterative BDD
(iBDD). We then propose a novel decoding algorithm that is based on
generalized minimum distance decoding of the component codes.
\RF{The proposed algorithm closes over $50$\% of the performance gap
between iBDD and turbo product decoding (TPD) based on the
Chase--Pyndiah algorithm. Moreover, the algorithm only leads to a
limited increase in complexity with respect to iBDD and has
significantly lower complexity than TPD.} The studied algorithms are
particularly interesting for high-throughput fiber-optic
communications.
\end{abstract}


\IEEEpeerreviewmaketitle

\section{Introduction}


The advent of codes-on-graphs and advances in digital electronics have
spurred a great deal of research on soft-decision forward error
correction (SD-FEC) for fiber-optic communications in the last decade,
see, e.g.,\cite{Lev14,Sch15,Hag15,Ksch17,Ksch18} and references
therein. Contrary to other applications such as wireless
communications---where SD-FEC is the de-facto standard---research on
SD-FEC in fiber-optic communications has been paralleled by a revival
of research on hard-decision FEC (HD-FEC). The reason is that SD-FEC
entails a significantly higher decoding complexity and data flow
compared to HD-FEC. Thus, for applications where very high throughputs
and low power consumption are required, HD-FEC is still an
appealing alternative. HD-FEC can also be combined with multilevel
modulation formats to achieve high spectral efficiency
\cite{Smith2012, Haeger2016istc, Sheikh2018}. 


Recent research on HD-FEC for fiber-optic communications has been
largely fuelled by the proposal of several new classes of product-like
codes, such as staircase codes \cite{staircase_frank} and braided
codes \cite{Feltstrom2009,Jia13}, which we refer to as generalized
product codes (GPCs). Similar to the original product codes (PCs)
\cite{Elias1954}, GPCs are built from smaller component codes,
typically Reed--Solomon or Bose--Chaudhuri--Hocquenghem (BCH) codes,
which can be efficiently decoded via algebraic bounded distance
decoding (BDD). The overall GPC can then be decoded by iteratively
applying BDD to the component codes. This algorithm is referred to as
iterative BDD (iBDD) and achieves an excellent performance--complexity
trade-off. 



GPCs can also be decoded iteratively by employing soft-input
soft-output (SISO) component decoding. This is referred to as turbo product
decoding (TPD) and typically implemented in practice via the
Chase--Pyndiah algorithm \cite{Pyndiahetal}.\footnote{TPD can
also be based on other component decoders, e.g., the forward-backward
algorithm applied to the component code trellis. In this paper, we use the term TPD
to refer to the iterative SISO decoding based on the Chase--Pyndiah algorithm.} TPD yields
larger net coding gains than iBDD but it has a significantly higher
decoding complexity. In order to (roughly) quantify the complexity
increase, one may rely for example on decoder data flow considerations
\cite{staircase_frank} or compare existing implementations (e.g.,
\cite{Condo2018, Vsatsheet}) in terms of gate counts. Both approaches
reveal that the complexity and potential power consumption increases
by around $1$--$2$ orders of magnitude when switching from iBDD to
TPD.  Given that commercially available implementations of TPD already
consume around $8$ W to achieve a throughput of $100$ Gb/s
\cite{Vsatsheet}, it remains a significant challenge to scale the
larger net coding gains to even higher throughputs. On the other hand,
staircase decoders based on iBDD remain feasible for throughputs as
large as $1$ Tb/s \cite{Fougstedt2018b}. 





The recent years have seen an increasing interest in the research
community in closing the gap between the performance of HD-FEC and
SD-FEC, while keeping the decoding complexity low. An interesting line
of research is to concatenate an inner SD-FEC code, e.g., a
low-density parity-check (LDPC) code decoded via belief propagation,
with an outer staircase code \cite{Ksch17,Ksch18}. Another
alternative, investigated by the authors in \cite{Hag18} and
\cite{She18}, is to improve the performance of iBDD. In \cite{Hag18},
a new anchor decoding (AD) algorithm that exploits conflicts between
component codes in order to assess their reliabilities, even when no
channel reliability information is available, is proposed. The
algorithm in \cite{She18} improves performance by exploiting channel
reliabilities, while still only exchanging binary (hard-decision)
messages between component codes, similar to iBDD.\footnote{A similar
approach was analyzed in the context of low-complexity decoding
algorithms for LDPC codes in \cite{lechner2012analysis}.} 




In this paper, we study and compare several decoding algorithms for
GPCs based on algebraic decoding of the component codes. Our focus is
on PCs, but the considered algorithms can also be applied to GPCs such
as staircase and braided codes. We start by reviewing the two decoding
algorithms that were proposed in \cite{Hag18} and \cite{She18},
respectively. It is shown that for the considered scenario, both
algorithms offer sizable net coding gain improvements of $0.18$ dB and
$0.25$ dB, respectively, with only a small complexity increase
compared to iBDD. We then propose a novel iterative decoding algorithm
based on generalized minimum distance (GMD) decoding of the component
codes, which we refer to as iterative GMD decoding with scaled
reliability (iGMDD-SR). Using iGMDD-SR, a more significant coding gain
improvement of around $0.60$ dB can be achieved compared to iBDD. This
closes over $50$\% of the performance gap to TPD, while maintaining
significantly lower complexity. 


\emph{Notation:} We use boldface letters to denote vectors
$\boldsymbol{x}$ and matrices $\boldsymbol{X}=[x_{i,j}]$. The $i$-th
row and $j$-th column of $\boldsymbol{X}$ are
denoted by $\boldsymbol{X}_{i,:}$ and $\boldsymbol{X}_{:,j}$, respectively. 
$|a|$ denotes the absolute value of $a$, and
$\left\lfloor a \right\rfloor$ the maximum integer value less than or
equal to $a$. 
$\mathcal{N}(\mu ,\sigma^2)$ represents a Gaussian
distribution with mean $\mu$ and variance $\sigma^2$. The Hamming
distance between vectors $\a$ and $\b$ is denoted by $\ham(\a,\b)$.

\section{Preliminaries}
\label{sys_mod} 

\begin{figure}[!t]
	\centering
	\includegraphics[width=3.4cm]{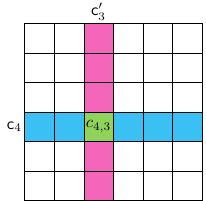}
	$\qquad$
	\includegraphics{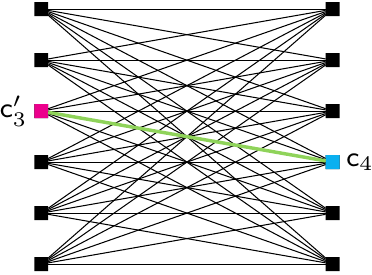}
	\caption{Code array (left) and simplified Tanner graph (right) for a PC assuming a
	component code of length $\nc=6$. In the simplified Tanner graph,
	degree-$2$ VNs are omitted and instead represented as simple edges. }
	\label{fig:PCSimpGraph}
	 \vspace{-2mm}
\end{figure}

Let $\mathcal{C}$ be a binary linear $(\nc, \kc, \dmin)$ code, where
$\nc$, $\kc$, and $\dmin$ are the code length, dimension, and minimum
distance, respectively. A (two-dimensional) PC with parameters
$(\nc^2,\kc^2,\dmin^2)$ and rate $R = \kc^2/\nc^2$ is defined as the
set of all $\nc\times\nc$ arrays such that each row and column of the
array is a codeword of $\mathcal{C}$. Accordingly, a codeword of the
product code can be represented as a binary matrix $\cc=[c_{i,j}]$ of
size $\nc^2\times \nc^2$. Alternatively, a PC can be defined via a
Tanner graph with $2\nc$ constraint nodes (CNs), where $\nc$ CNs
correspond to the row codes and $\nc$ CNs correspond to the column
codes. The graph has $\nc^2$ variable nodes (VNs) corresponding to the
$\nc^2$ code bits.  The code array and (simplified) Tanner graph of a
PC where $\nc=6$ is shown in Fig.~\ref{fig:PCSimpGraph}. 




We assume transmission over the binary-input additive white Gaussian
noise (bi-AWGN) channel. In particular, the
channel observation corresponding to code bit $c_{i,j}$ is given by
\begin{align}\label{channel_inst}
y_{i,j}=x_{i,j}+z_{i,j},
\end{align}  
where $x_{i,j}=(-1)^{c_{i,j}}$ and $z_{i,j}\sim \mathcal{N}(0,(2 R
E_\mathrm{b}/N_0)^{-1})$. Let $\lalone=[L_{i,j}]$ be the matrix of channel
log-likelihood ratios (LLRs) and $\rr=[r_{i,j}]$ the matrix of hard
decisions at the channel output, where $r_{i,j}$ is obtained by
mapping the sign of $L_{i,j}$ according to $- 1 \mapsto 0$, $+ 1
\mapsto 1$. This mapping is denoted by $\BB(\cdot)$, i.e.,
$r_{i,j}=\BB(L_{i,j})$. With some abuse of notation, we also write
$\rr=\BB(\lalone)$.

\subsection{Bounded Distance Decoding}
\label{BDD}



Consider now the decoding of an arbitrary row or column component
code, assuming that the codeword $\lc=(c_1,\ldots,c_{\nc})$ is
transmitted and only hard-detected channel observations
$\lr=(r_1,\ldots,r_{\nc})$ are available. BDD corrects all
error patterns with Hamming weight up to the error-correcting
capability $t = \left\lfloor \frac{\dmin-1}{2} \right\rfloor$ of the code. If the
weight of the error pattern is larger than $t$ and there exists
another codeword $\tilde{\lc} \in \mathcal{C}$ with
$\mathsf{d_H}(\tilde{\lc},\lr)\le t$, then BDD maps $\lr$ to
$\tilde{\lc}$ and thus introduces a miscorrection.  Otherwise, if no
such codeword exists, BDD fails and we use the convention that the
decoder outputs $\lr$. Thus, the decoded vector $\hat{\rbol}$ for BDD
can be written as
\begin{equation}
	\label{eq:BDD_VN}
	\hat{\rbol} = \begin{cases}
		\lc & \text{if } \ham(\lc, \rbol)  \leq t \\
		\tilde{\lc} \in \mathcal{C} & \text{if } \ham(\lc, \rbol) > t \text{ and
		} \ham(\tilde{\lc}, \rbol) \leq t \\
		\rbol & \text{otherwise}
	\end{cases}.
\end{equation}

\subsection{Generalized Minimum Distance Decoding}
\label{sec:GMD}





Consider again the component decoding but now assume that the vector
of channel LLRs $\llr=(L_1,\ldots,L_{\nc})$ is available. In that
case, GMD decoding can be employed which is based on multiple
algebraic error-erasure decoding attempts \cite{ForneyGMD}.  In
particular, the decoder ranks the coded bits in terms of their
reliabilities $|L_1|,\ldots,|L_{\nc}|$. Then, the $m$ least reliable
bits in $\lr$ are erased, where $m \in \{\dmin-1,\dmin-3,...,2\}$ if
$\dmin$ is odd and $m \in \{\dmin-1,\dmin-3,...,3\}$ if $\dmin$ is
even. Together with $\lr$, this gives a list of $t+1$ trial vectors
$\lrt_{i}$, out of which $t$ vectors contain both erasures and
(possibly) errors. Finally, algebraic error-erasure decoding
\cite[Sec.~6.6]{LinCos04} is applied to each $\lrt_i$, $i = 1, \dots,
t+1$.
If error-erasure decoding fails for all $t+1$ vectors in the list, an
overall failure is declared for the GMD decoding. On the other hand,
if some of the error-erasure decoding attempts did not fail, the
decoder picks among all decoded candidate codewords $\hat{\lr}$ the
one that minimizes the generalized distance \cite{ForneyGMD}
\begin{align}\label{GMDmetric}
\dGMD=\sum\limits_{i:{r_{i}} = {\hat{r}_{i}}} {\left( {1 - {\alpha_{i}}} \right)}  + \sum\limits_{i:{r_{i}} \ne {\hat{r}_{i}}} {\left( {1 + {\alpha_{i}}} \right)}, 
\end{align}
where $\alpha_{i}\buildrel \Delta \over = |L_{i}|/\mathop {\max
}\limits_{1 \le j \le \nc} |L_{j}|$. Note that if all input LLRs
$L_{j}$ have the same magnitude, we have $\alpha_i = 1$ for all $i =
1,\dots,\nc$ and \eqref{GMDmetric} reverts to $2 \ham(\lr, \hat{\lr})$. 



\section{Iterative Decoding of Product Codes}

PCs can be decoded by applying BDD of the row and column codes in an
iterative manner. This algorithm is referred to as iBDD. In the
following, two other algorithms that can improve the
performance of iBDD are described. 


\subsection{Anchor Decoding}

When a miscorrection occurs during iBDD, it is possible that two
component codes disagree on the value of a particular bit, leading to
a conflict. Conflicts are typically ignored in the sense that row and
column codes are decoded sequentially and previous decoding decisions
are simply overridden. The main idea in AD is to introduce status
information for each component code and designate certain ``reliable''
component codes as anchors. Then, no further additional corrections
from other component codes are allowed if this would lead to a
conflict and overturn the decision of an anchor. Since some anchors
may actually be miscorrected, AD also allows for the backtracking of
the decoding decisions of anchors. This happens whenever too many
other component codes are in conflict with a particular anchor.
Pseudocode for AD can be found in \cite[Alg.~2]{Hag18}. 

\subsection{Iterative Bounded Distance Decoding With Scaled Reliability}
\label{BDD-SRsec}

Both iBDD and AD do not take potentially available channel reliability
information into account. In \cite{She18}, we proposed a modification
of iBDD where channel reliabilities are exploited, while only
binary messages between component decoders are exchanged. This algorithm is
referred to as iBDD with scaled reliability (iBDD-SR). In particular,
assume that the $i$-th row code has been decoded via BDD. In order to
combine the BDD output with the channel LLRs, the decoded bits are
mapped according to $0 \to +1$ and $1 \to -1$ if BDD is successful and
mapped to $0$ if a decoding failure occurs. Let
$\bar{\mu}_{i,j}^{\mathsf r, (\ell)} \in \{\pm1, 0 \}$ be the result
of this mapping for the decoded bit corresponding to code bit $c_{i,j}$ in iteration $\ell$. Then,
we compute
\begin{equation}\label{eq:BDDchrel_VN_scale}
\psi_{i,j}^{\mathsf{r},(\ell)}=
\BB(\w_\ell \cdot \bar{\mu}_{i,j}^{\mathsf r, (\ell)} + L_{i,j}),
\end{equation}
where $w_\ell>0$ is a scaling parameter and
$\psi_{i,j}^{\mathsf{r},(\ell)}$ can be interpreted as the message
passed from the $i$-th row code to the $j$-th column code. In
particular, after applying this procedure to all row codes, the matrix
$\boldsymbol{\Psi}^{\mathsf{r},(\ell)}=[\psi_{i,j}^{\mathsf{r},(\ell)}]$
is used as the input for the column codes, where BDD based on
$\boldsymbol{\Psi}^{\mathsf{r},(\ell)}$ is performed. The binary
output messages for the row codes are then formed in a similar fashion
as for the column codes. Intuitively, the mapping
\eqref{eq:BDDchrel_VN_scale} helps to alleviate the effect of
miscorrections by allowing the outcome of BDD at certain bit positions
to be overturned if the corresponding channel LLR is very reliable
(i.e., $|L_{i,j}|$ is large).






\section{Iterative Generalized Minimum Distance Decoding with Scaled Reliability}\label{GMD}

In this section, we propose a novel iterative decoding algorithm for
PCs based on GMD decoding of the component codes and the exchange of
soft information between component codes. We refer to this algorithm
as iGMDD-SR.  


iGMDD-SR works as follows. Without loss of generality, assume that the
decoding starts with the row codes and let us consider the decoding of
the $i$-th row code at iteration $\ell$.  Let
$\tilde{\boldsymbol{\mu}}_i^{\mathsf{r},(\ell)}$ be the vector of soft
information corresponding to $\cc_{i,:}$ at the input of the $i$-th
row decoder at iteration $\ell$, resulting from the decoding of the
$\nc$ column codes at decoding iteration $\ell-1$, where, initially,
$\tilde{\boldsymbol{\mu}}_i^{\mathsf{r},(1)} = \lalone_{i,:}$. Also,
let
$\rr^{(\ell)}_{i,:}=\BB(\tilde{\boldsymbol{\mu}}_i^{\mathsf{r},(\ell)})$
be the corresponding hard-decoded vector. Then, GMD decoding of the
$i$-th row code is performed as explained in Section~\ref{sec:GMD}
based on $\rr^{(\ell)}_{i,:}$ and the reliabilities
$|\tilde{\boldsymbol{\mu}}_i^{\mathsf{r},(\ell)}|$. Note that GMD
decoding does not provide reliability information about the decoded
bits, i.e., it is ``soft-input, hard-output''. In order to provide
reliability information to the column decoders, we resort to a
heuristic approach that is similar to the approach used for iBDD-SR.
In particular, the output bits of GMD decoding are mapped according to
$0 \to +1$ and $1 \to -1$ if GMD decoding is successful and mapped to $0$
if GMD decoding fails. Let $\bar{\mu}_{i,j}^{\mathsf r, (\ell)} \in
\{\pm1, 0 \}$ be the result of this mapping for the decoded bit corresponding to code bit $c_{i,j}$. The reliability information is then formed
according to
\begin{equation}\label{eq:GMDchrel_VN_scale}
\mu_{i,j}^{\mathsf r, (\ell)}=\w_\ell \cdot \bar{\mu}_{i,j}^{\mathsf r,
(\ell)} + L_{i,j}, 
\end{equation}
where $\w_\ell > 0$ is a scaling parameter and
$\boldsymbol{\mu}_i^{\mathsf{r},(\ell)}=(\mu_{i,1}^{\mathsf{r},(\ell)},\ldots,\mu_{i,\nc}^{\mathsf{r},(\ell)})$
is the entire soft-output vector of the $i$-th row decoder. 

\begin{figure}[t!]
	\centering
	\includegraphics[width=\columnwidth]{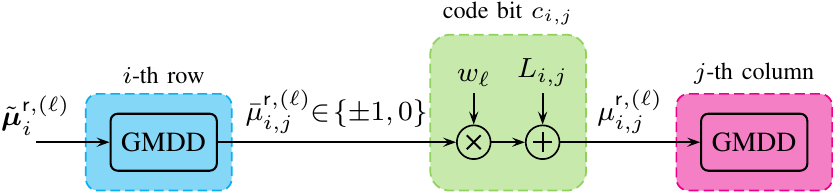}
	\caption{Block diagram showing the information flow from the $i$-th row to
	the $j$-th column code in iGMDD-SR.}
	\label{messagepasing2}
\end{figure}

Visualizing the decoding over the Tanner graph of the code,
$\mu_{i,j}^{\mathsf{r},(\ell)}$ corresponds to the message from row CN
$i$ to column CN $j$. Now assume that all row codes have been decoded.  The
vector of soft information corresponding to $\cc_{:,j}$ at the input
of the $j$-th column decoder is then defined as 
$\tilde{\boldsymbol{\mu}}_j^{{\mathsf
c},(\ell)}=(\tilde{\mu}_{j,1}^{\mathsf{c},(\ell)},\ldots,\tilde{\mu}_{j,\nc}^{\mathsf{c},(\ell)})=(\mu_{1,j}^{\mathsf{r},(\ell)},\ldots,\mu_{\nc,j}^{\mathsf{r},(\ell)})$.
GMD decoding is performed based on
$\rr^{(\ell)}_{:,j}=\BB(\tilde{\boldsymbol{\mu}}_j^{{\mathsf
c},(\ell)})$ and $|\tilde{\boldsymbol{\mu}}_j^{{\mathsf c},(\ell)}|$.
The soft output of the $j$-th column decoder is formed similar to
\eqref{eq:GMDchrel_VN_scale} and the resulting soft output vector is
denoted by $\boldsymbol{\mu}_j^{{\mathsf
c},(\ell)}=(\mu_{j,1}^{\mathsf{c},(\ell)},\ldots,\mu_{j,\nc}^{\mathsf{c},(\ell)})$,
where $\mu_{j,i}^{{\mathsf c},\ell}$ corresponds to the message from
column CN $j$ to row CN $i$. After decoding all column component
codes, we set
$\tilde{\boldsymbol{\mu}}_i^{\mathsf{r},(\ell+1)}=(\tilde{\mu}_{i,1}^{\mathsf{r},(\ell+1)},\ldots,\tilde{\mu}_{i,\nc}^{\mathsf{r},(\ell+1)})=(\mu_{1,i}^{\mathsf{c},(\ell)},\ldots,\mu_{\nc,i}^{\mathsf{c},(\ell)})$
and the iterative process continues until a maximum number of
iterations is reached. The information flow from
the row to column codes in iGMDD-SR is schematically illustrated in
Fig.~\ref{messagepasing2}.




\section{Decoding Complexity Discussion}
\label{complexity}



A thorough complexity analysis of the different decoding algorithms
presented in Sections~\ref{BDD}, \ref{BDD-SRsec}, and \ref{GMD} is a
formidable task that should include, besides the pure algorithmic
aspects, also the implications in terms of internal data flow. A
complete analysis is therefore out of the scope of this paper. We
however provide a high-level discussion of complexity aspects
associated with the different choices of the component code decoders,
focusing on the complexity per decoding iteration. Additional remarks
on the decoding complexity of AD and iBDD-SR can also be found in the
respective papers \cite{Hag18} and \cite{She18}.

The use of BDD to decode the component codes represents the simplest
approach among those considered. From this viewpoint, iBDD, AD, and
iBDD-SR are characterized by a similar complexity. For iBDD-SR, the
combination of the BDD output and the channel LLRs in
\eqref{eq:BDDchrel_VN_scale} yields a small complexity increase with
respect to iBDD and AD. With respect to BDD of the component codes,
GMD decoding entails more substantial changes in the decoder.  In this
case, the component decoder has to be provided with soft decisions by
the previous decoding step. Finally, $t$ error-erasure decoding attempts and one BDD attempt
are required. Each error-erasure decoding attempt  has a cost  close to a run of BDD. Each
decoding attempt may result in a candidate codeword that is used to
form a list of size up to $t+1$. The minimization of the distance in
\eqref{GMDmetric} has a negligible cost with respect to the $t+1$
decoding attempts.


The decoding complexity of TPD exceeds the complexity of AD, iBDD-SR,
and iGMDD-SR. In particular, the Chase--Pyndiah algorithm requires the
construction of a list of binary test sequences for each component
decoding, where the list size depends on a design parameter.
Typically, the list size is set to at least $16$ \cite{Pyndiahetal},
and BDD is applied to each test sequence. Thus, for components codes
where $t$ is small (e.g., $2$, $3$, or $4$), as the ones considered in
this paper and of practical use for high-throughput fiber-optic
communications, the list size and corresponding cost per component
decoding in GMD decoding is only a fraction of that of TPD. Moreover,
iGMDD-SR also relaxes the computational requirements when computing
the extrinsic soft-output information for each code bit compared to
TPD by using the heuristic update equation in
\eqref{eq:GMDchrel_VN_scale}.





\newcommand{\tablehighlight}{}
\begin{table*}[t]
 \caption{Comparison of different product decoding algorithms. Coding
	gains and capacity gaps are measured at $\text{BER} = 10^{-5}$.}
  \centering
  \renewcommand{\arraystretch}{1.2}
  \begin{tabular}{cccccc}
  		\toprule
		\tablehighlight{acronym} & \tablehighlight{decoding algorithm} &
		\makecell{\tablehighlight{channel}\\ \tablehighlight{reliabilities}} &
		\makecell{\tablehighlight{exchanged} \\
		\tablehighlight{messages}} & \makecell{gain over\\ iBDD [dB]}&
		\makecell{gap from\\ capacity [dB]}\\
		\midrule
		iBDD & iterative bounded distance decoding & no & hard & - & $0.98$ (HD) \\
		AD & anchor decoding \cite{Hag18} & no & hard & $0.18$ & $0.80$ (HD) \\
		iBDD-SR & iterative bounded distance decoding with scaled
		reliability \cite{She18} & yes & hard  & $0.25$ & $2.00$ (SD) \\
		iBDD (ideal) & iterative bounded distance decoding without
		miscorrections & no & hard & $0.28$ & $0.70$ (HD) \\
		iGMDD-SR & iterative generalized minimum distance decoding with
		scaled reliability & yes & soft & $0.60$ & $1.66$ (SD) \\
		TPD & turbo product decoding (Chase--Pyndiah) \cite{Pyndiahetal} & yes & soft & $1.08$ & $1.18$ (SD) \\
		TPD (Viasat) & commercially available decoder (undisclosed
		component code details) \cite{Vsatsheet} & yes & soft & $1.26$ & $1.00$ (SD) \\
		\bottomrule
  \end{tabular}
\end{table*}

\section{Simulation results}

\renewcommand{\w}{\boldsymbol{w}}
\newcommand{\maxIter}{\ell_{\text{max}}}


In this section, we compare the product decoding algorithms in terms
of their performance. For the simulations, we consider
double-error-correcting extended BCH (eBCH) codes with parameters
$(256,239,6)$ as component codes. The resulting PC has rate 
$R = 239^2/256^2 \approx 0.8622$ corresponding to an FEC overhead
of $1/R-1 \approx 15 \%$. For all algorithms, a maximum of $\maxIter =
10$ decoding iterations is performed. 

Both iBDD-SR and iGMDD-SR require a proper choice for the scaling
factors $w_\ell$ in each iteration. We jointly optimize all scaling
factors $\w = (w_1, \dots, w_{\maxIter})$ by using Monte--Carlo
estimates of the bit error rate (BER) for a fixed $E_\mathrm{b}/N_0$ as the
optimization criterion. Intuitively, one would expect that the
decisions of the component decoders become more reliable with
iterations, whereas the channel observations become less informative.
Therefore, in order to reduce the optimization search space, we only
consider vectors $\w$ with monotonically increasing entries.
For iBDD-SR, we found that the optimized vector $\w$ is relatively
insensitive to the targeted $E_\mathrm{b}/N_0$.
On the other hand, iGMDD-SR is more
sensitive to a mismatch between the optimized and actual
$E_\mathrm{b}/N_0$ and the optimization is thus performed for each
value of $E_\mathrm{b}/N_0$ separately. 




\begin{figure}[t] \centering 
	
	\includegraphics[width=\columnwidth]{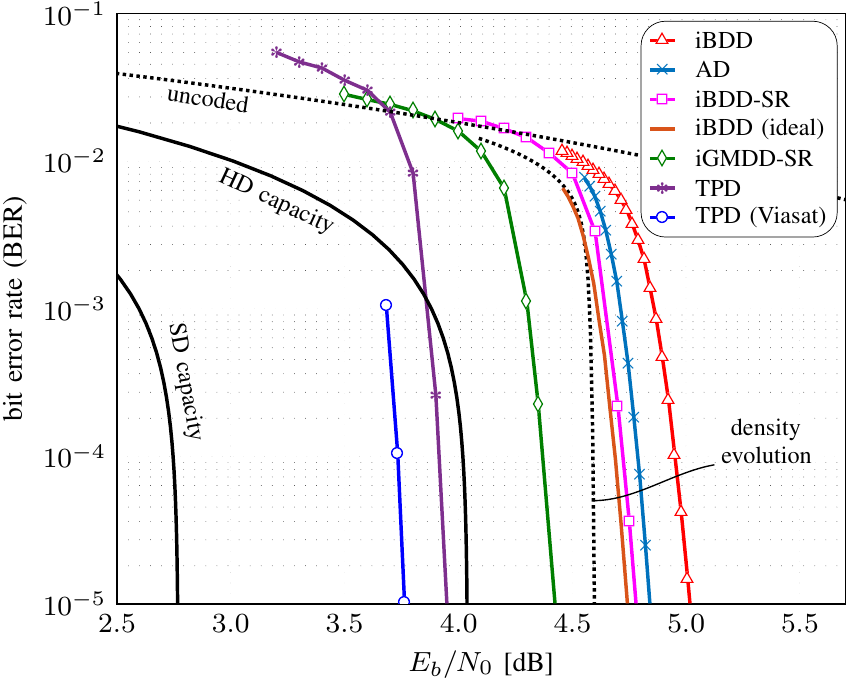}  
	\caption{Performance of different product decoding algorithms
	for ($256$,$239,6$) eBCH component codes and $10$ iterations. The PC rate is $0.8622$
	corresponding to $15 \%$ FEC overhead. For TPD (Viasat), results are taken
	from \cite{Vsatsheet} for the same overhead, but different component codes may be employed.} 
	\label{v9t4s7_product} 
\end{figure}

In Fig.~\ref{v9t4s7_product}, the performance of iBDD, AD, iGMDD-SR,
and iBDD-SR is shown. We also plot the performance of TPD via
off-the-shelf Matlab toolbox functions and compare to a commercially
available $100$ Gb/s SD-FEC solution implementing TPD for the same
overhead \cite{Vsatsheet}. The data points are directly extracted from
\cite{Vsatsheet}, but we remark that the component code details are
not disclosed in \cite{Vsatsheet}. Thus, a different PC may be used.
Moreover, pre- and post-processing steps are employed, which may also
explain the performance difference.

AD and iBDD-SR outperform the conventional iBDD by $0.18$ dB and
$0.25$ dB at a bit error rate (BER) of $10^{-5}$, respectively. As a
reference, we show the performance of idealized iBDD, where a genie
prevents all miscorrections. The asymptotic performance of idealized
iBDD can be analytically predicted by using density evolution
\cite{Justesen2011, Haeger2017tit}, which is shown by the dotted line.
It can be seen that both AD and iBDD-SR are effective algorithms to
combat miscorrections. The performance degradation of iBDD-SR compared
to ideal iBDD is very small ($< 0.01$ dB), implying that by properly
tuning the scaling parameters $\w$, iBDD-SR can alleviate the effect
of miscorrections to a large extent. 

One can also see that iGMDD-SR outperforms iBDD, AD, and iBDD-SR. In
particular, the performance gain of iGMDD-SR over iBDD is $0.60$ dB at
a BER of $10^{-5}$. The additional performance gain is expected, since
GMD decoding can decode beyond half the minimum distance by
introducing erasures and performing multiple error-erasure component
decoding attempts. Furthermore, iGMDD-SR performs $0.52$ dB away from
TPD, i.e., it closes over $50$\% of the performance gap between iBDD
and TPD.

The net coding gain improvements of all considered decoding algorithms
over iBDD are summarized in Table I. We also indicate the gap to
capacity for all schemes. Note that the performance of iBDD and AD
should be compared to the HD capacity, whereas the performance of
iBDD-SR, iGMDD-SR, and TPD should be compared to the SD capacity since
channel LLRs are exploited during decoding. Overall, one can see a
clear trade-off between performance and complexity for the different
algorithms, e.g., using iGMDD-SR, with higher complexity than iBDD,
yet less complexity than TPD, the gap between iBDD and TPD is
approximately halved.

\section{Conclusion}
\label{conclusion}

We studied several low-complexity iterative decoding algorithms for
PCs that outperform the conventional iBDD. In particular, we reviewed
two previously proposed algorithms, AD and iBDD-SR, and we proposed a
novel algorithm called iterative GMD decoding with scaled reliability
(iGMDD-SR). For the considered scenario based on
double-error-correcting eBCH component codes, AD and iBDD-SR
outperform iBDD by $0.18$ dB and $0.25$ dB, respectively, with only a
small increase in complexity. The complexity increase for iGMDD-SR is
larger, but the algorithm achieves a more significant performance gain
of $0.60$ dB over iBDD. This closes over $50$\% of the performance gap
to TPD, at a significantly lower complexity.


\end{document}